\documentclass{article}

\usepackage{arxiv}

\usepackage[utf8]{inputenc} % allow utf-8 input
\usepackage[T1]{fontenc}    % use 8-bit T1 fonts
\usepackage{hyperref}       % hyperlinks
\usepackage{url}            % simple URL typesetting
\usepackage{booktabs}       % professional-quality tables
\usepackage{amsfonts}       % blackboard math symbols
\usepackage{nicefrac}       % compact symbols for 1/2, etc.
\usepackage{microtype}      % microtypography
\usepackage{graphicx}
\usepackage{amssymb}
\usepackage{amsmath}
\usepackage{lineno}
\usepackage{hyperref}
\usepackage{listings}
\usepackage{subfigure}
\usepackage{comment}
\usepackage{enumitem}
\usepackage[dvipsnames]{xcolor}

\title{Deep Learning-based extraction of surface wave dispersion curves from seismic shot gathers}

\author{
  Danilo Chamorro\\
  KAUST\\
  Thuwal, Kingdom of Saudi Arabia\\
  \texttt{dchamorror@kaust.edu.sa}\\
  \And
  Jiahua Zhao\\
  University of Padua\\
  Padua, Italy\\
  \texttt{jiahua.zhao@studenti.unipd.it}
    \And
  Claire Birnie\\
  KAUST\\
  Thuwal, Kingdom of Saudi Arabia\\
  \texttt{claire.birnie@kaust.edu.sa}
    \And
  Myrna Staring\\
  Fugro\\
  Delft, Netherlands\\
  \texttt{m.staring@fugro.com}
    \And
  Moritz Fliedner\\
  Fugro\\
  Delft, Netherlands\\
  \texttt{M.Fliedner@fugro.com}
    \And
  Matteo Ravasi\\
  KAUST\\
  Thuwal, Kingdom of Saudi Arabia\\
  \texttt{matteo.ravasi@kaust.edu.sa}
  }
  
\begin{document}

\chead{Deep Learning-based extraction of surface waves}

\maketitle

\begin{abstract}
   Multi-channel Analysis of Surface Waves (MASW) is a seismic method employed to obtain useful information about shear-wave velocities in the near surface. A fundamental step in this methodology is the extraction of dispersion curves from dispersion spectra, which are obtained after applying specific processing algorithms onto the recorded shot gathers. Whilst this extraction process can be automated to some extent, it usually requires extensive quality control, which can be arduous for large datasets. We present a novel approach that leverages deep learning to identify a direct mapping between seismic shot gathers and their associated dispersion curves (for both fundamental and first modes), by-passing therefore the need to compute dispersion spectra. Given a site of interest, a set of 1D velocity and density models are created using prior knowledge of the local geology; pairs of seismic shot gathers and Rayleigh-wave phase dispersion curves are then numerically modeled and used to train a simplified residual network. The proposed approach is shown to achieve high quality predictions of dispersion curves on a synthetic test dataset and is, ultimately, successfully deployed on a field dataset. Various uncertainty quantification and CNN visualization techniques are also developed to assess the quality of the inference process and better understand the underlying learning process of the network. The predicted dispersion curves are finally inverted, and the resulting shear-wave velocity model is shown to be plausible and consistent with prior geological knowledge of the area.
\end{abstract}

\section{Introduction}
Multi-channel Analysis of Surface Waves (MASW) is a well-established seismic method commonly employed to obtain information about the near surface, specifically its shear-wave velocities, using the measured dispersive behavior of surface waves \cite{Moro2015, Park2007}. The MASW method is able to discriminate the multimodal Rayleigh wave phenomena from other types of both surface and body waves, and thus its propagation properties are highly related to the site conditions \cite{Penumadu2005}.  
A fundamental processing step in the MASW methodology is represented by the extraction of dispersion curves from dispersion spectra. These algorithms transform the shot gathers into an alternative domain, relating phase velocity versus frequency to illustrate the dispersion of surface waves in the subsurface. Various algorithms can be used for this purpose, with most of them being linearly dependent on each other and differing mainly in terms of noise sensitivity, robustness, and the weighting of information \cite{Socco2004}. Because of its robustness to such factors, the approach of  Park \cite{Park2007} is the one most generally used. A major restriction of this technique arises from the lack of spectral resolution as a result of the acquisition limitations; this creates a windowing effect over the dispersion spectra, which ultimately makes the process of dispersion curve picking difficult and prone to interpretation errors \cite{Socco2004}. Ultimately, once dispersion curves are extracted, MASW can produce a 2-D Vs model of the subsurface by performing a dispersion inversion in an efficient and robust manner \cite{Penumadu2005}; however, the accuracy of the estimated model is highly dependent on the quality of the dispersion curves picking \cite{Baglari2018}. 
Although the dispersion curve picking process can be automated to some extent, it requires quality control by experts, which can be time-consuming and inevitably affected by subjectivity - becoming a bottleneck when dealing with large datasets. To overcome such limitations, several methodologies have been proposed in the literature: \cite{Zheng2014} developed an automatic workflow using binarization and thinning over the obtained dispersion spectra; similarly, \cite{Taipodia2020} presented an automatic picking image processing algorithm based on a manually specified energy threshold. As a promising alternative to the traditional methods, the use of machine learning techniques for dispersion curve picking has also been suggested by several authors: \cite{Colombo2019} developed a deep belief network to pick fundamental modes in the phase velocity spectrum, however, this approach was limited in the training side since the dataset used for this was picked manually over dispersion spectra panels, requiring big amounts of effort for a field-centric approach as the one we propose; alternatively, \cite{Rovetta2020} used a density-based spatial clustering algorithm for automatic picking of surface wave dispersion curves by training over synthetic dataset, showing promising results while avoiding the required training stage of the machine learning approaches; finally, \cite{Kaul2021} proposed a deep learning model using a fully convolutional architecture with residual units to pick multi-mode dispersion curves through pixel-wise binary segmentation from dispersion spectra.
Most of these approaches work under the assumption that the medium is normally dispersive, has no velocity inversions, and has no strong velocity contrasts. Therefore, the fundamental mode is energy dominant, and the picking of the dispersion curve is related to picking the maximum energy over the phase-velocity spectra \cite{Socco2004}. This assumption is valid in most cases given the presence of simply layered subsurface; however, a visually continuous energy peak may shift between modes at certain frequencies \cite{Pan2019}, an effect known as mode osculation. This usually happens in the presence of strong velocity contrasts, typically in places where a fast bedrock underlies loose sediments \cite{Boaga2013}. Another common limitation of these approaches is that the information from the dispersion spectra is being used and not the one from the seismic shot gathers themselves; this can be counterproductive in the sense that there are many methods for obtaining such spectra; for example, when the technique developed by \cite{Park2007} is used, the maximum values in the dispersion panels do not always correlate with the fundamental mode \cite{Moro2015}. Moreover, the quality of the dispersion spectra heavily depends on the spatial sampling of the receiver array, leading to a non-unique mapping between them and their underlying dispersion curves. Consequently, the output of such automatic methods may be biased, something that may affect the subsequent inversion results \cite{Zhang2003}. 
Our work presents a novel methodology for estimating multi-mode dispersion curves directly from seismic shot gathers, utilizing a Convolutional Neural Network guided by geophysical theory. Firstly, a set of 1D velocity and density models is created for a given site of interest, based on prior geological knowledge of the near surface. Next, the corresponding seismic shot gathers are numerically modelled alongside their associated Rayleigh-wave phase dispersion curves, and these are used to train a simplified residual network. The network learns the direct mapping from the shot gathers to their corresponding dispersion curves. Our proposed approach achieves satisfactory predictions of dispersion curves on a synthetic test dataset and is later successfully applied to a field dataset. Two approaches to uncertainty quantification (referred to as UQ) in neural network predictions, namely Monte-Carlo dropout and Distributional Parameter Estimation, are also investigated as a way to provide confidence intervals alongside the predicted dispersion curves. This is valuable to assess the quality of the inference process and can also be used as additional input to the subsequent inversion process. Moreover, we have experimented with a technique from the field of Explainable AI (referred to as XAI) to obtain a deeper understanding of the network’s learning process: more specifically, an auxiliary output can be created to identify which areas within the input dataset contribute to the prediction of different parts of a dispersion curve. Finally, the predicted dispersion curves are inverted to create a 2D shear-wave velocity model of the subsurface. Our results are plausible and consistent with prior geological knowledge of the area of interest.

\section{Methodology}

In this section, we detail the overall workflow proposed in this work to predict dispersion curves from synthetic seismic shot gathers as schematically represented in Figure 1. Special attention is devoted to the implementation details that allow our model to generalize to field data.

\subsection*{Data generation}

To successfully train a supervised machine learning algorithm, a labeled training dataset is required in order to train a model; moreover, a subset of the training data is commonly reserved for validation purposes to be able to assess the model's prediction capabilities. In our particular scenario, we need to create a collection of dispersion curves and corresponding seismic shot gathers that encompass a diverse range of geological conditions, while also considering the general geology of the site where the model will operate. This involves determining an estimate of the possible number of layers, velocity ranges, and main acquisition parameters of the survey, so that we can select a suitable parametrization for the 1-D velocity and density profiles. This process enables us to randomly generate 1000 pairs of synthetic shot gathers and dispersion curves that conform to the anticipated distribution. The synthetic shot gathers are numerically modelled using an elastic finite-difference modeling engine \cite{Virieux2012}, whilst the Dunkin’s matrix method \cite{Dunkin1965} is used to create the dispersion curves. In our specific work, this is accomplished with the open-source tools Fdelmodc \cite{Thorbecke2011} and Disba \cite{Luu2021}, respectively. In our experiments, both the seismic dataset and the dispersion curves are modelled using a flat wavelet in the frequency range between 5 Hz to 60 Hz; this is to permit our training to be done on a less wavelet-dependent domain while also limiting the frequency to a physically expected range. 
Finally, we must note that we take a field-centric approach, meaning that no one model is expected to be able to predict dispersion curves from shot gathers for any geological scenario and geophysical acquisition setup; conversely, the training data and, therefore, the trained model are specialized to one specific setting making the synthetic data generation process highly coupled with the field data we wish to apply the model to at later stage. In our study, we allow a velocity variation of 35\% from the initial velocity model, and a variation in the number of layers between 2 and 4. These ranges enable us to accommodate for the errors and uncertainties that are naturally present in our prior knowledge of the geological setting.

\subsection*{Data pre-processing}

Considering inevitable differences in the seismic signal (and noise) between clean synthetic and field data, a data pre-processing sequence is introduced: the field data is initially low-pass filtered, and an average statistical wavelet is estimated from the entire dataset. This wavelet is then used to shape the spectrum of the synthetic dataset such that it more closely resembles that of the field data used at the inference stage. Similarly, the field dataset is normalized and convolved with the wavelet used to create the training dataset. Finally, realistic colored noise is also added to the modeled data. 
Moreover, noting that the field data exhibits some strong linear events at early times that cannot be reproduced by our finite-difference modelling sequence, random linear events are also added to the synthetic data: by adding such events to the synthetic data, we observe increased robustness of the trained model once applied to the field data at inference stage.

\subsection*{Training process}

The process of predicting dispersion curves from seismic shot gathers can be interpreted as a domain translation process between a 2-dimensional array (i.e., an image) and a 1-dimensional array (i.e., a ‘time’ series). Convolutional neural networks with a series of convolutional layers followed by one or more dense layers are commonly used for such a task. In this work, we choose a deep residual convolutional neural network; more specifically, we use a lightweight ResNet architecture, namely ResNet-18, which contains a relatively small number of trainable parameters (in comparison to other popular ResNet architectures \cite{He2015}). Minor modifications are applied to the original architecture to account for differences in the dimensions of our inputs and outputs: the input channel of the first layer is adjusted to a single channel given that our shot gathers are from a single-component seismic acquisition system, whilst the output dimension of the last dense layer is modified to produce an output whose size corresponds to that of the frequency axis of the dispersion curve of interest. In our numerical examples, an array of 500 samples is used for the target. Dropout with a probability of 20\% is finally added after every Rectified Linear Unit (ReLU) activation function to regularize the learning process and reduce overfitting to the training data; this strategy proved successful as predictions on the validation set show clear improvements over those produced by the same network without dropout. In order to train the network, the Huber norm between the true and predicted dispersion curves is chosen as the primary loss for the network training; a secondary loss is also introduced to enhance the smoothness of the predicted curves. The overall loss can be written as:

\begin{equation*}
    J = \frac{1}{N_s} \sum^{N_s}_{i=1} \mathcal{L}_H \left( y^{(i)}, f_\theta(x^{(i)}) \right) + \alpha || D f_\theta (x^{(i)}) ||^2_2
\end{equation*}

where $N_s$ is the number of training samples, $x^{(i)}$ is the i-th input shot gathers, $y^{(i)}$ is the i-th target dispersion curve, $\mathcal{L}_H$ is the Huber loss, $D$ is a second-order derivative operator used to penalize roughness in the predictions (i.e., encourage smoothness in the dispersion curves) and $\alpha$ is the regularization parameter. In all our experiments, a randomly selected portion of 20\% of the generated synthetic pairs of shot gathers and dispersion curves is used as the validation dataset. As for the training process, the Adam optimizer with a learning rate of 0.001 is used in this work. Moreover, an early stopping checkpoint with a patience parameter of 128 epochs is employed to monitor whether the validation loss is decreasing through epochs. Training is automatically stopped after 1000 epochs if the early stopping criteria has not been met.

\subsection*{Multi-mode prediction}

The proposed methodology is currently limited to predicting one single model (e.g. fundamental mode); however, it is well known that using multiple modes in inversion can reduce the ambiguity associated with the dispersion curve-to-velocity mapping.  We aim here to extend our methodology to allow for the simultaneous prediction of multiple modes. By doing so, we also aim to further constrain the results of the inference process and teach the network to predict more features exploiting the same information. For this approach, both first and fundamental dispersion curves were generated using the Disba package introduced previously; taking into account that the first mode has a smaller frequency range, the dispersion curve for the first mode was generated for frequencies from 15 Hz up to 60 Hz, while the fundamental mode frequency range was maintained the same as in the previous example. 

For the case of multi-mode prediction, an extended array of 1000 samples is used as the output dimension of the network. We define  $y^{(i)}=[y_0^{\left(i\right)},y_1^{(i)}]$ where $y_0^{\left(i\right)}$ and $y_1^{\left(i\right)}$ are the fundamental and the first mode dispersion curves, respectively. Similarly we write $f_\theta\left(x^{(i)}\right)=[\ f_\theta\left(x^{\left(i\right)}\right)_0,\ f_\theta\left(x^{\left(i\right)}\right)_1]$ where $f_\theta\left(x^{\left(i\right)}\right)_0$ and $f_\theta\left(x^{\left(i\right)}\right)_1$ are the predicted fundamental and the first mode dispersion curves, respectively. In this case, two regularization terms are introduced to encourage both predictions to be smooth and avoid artifacts at the inference stage:

\begin{equation*}
    J = \frac{1}{N_s} \sum^{N_s}_{i=1} \mathcal{L}_H \left( y^{(i)}, f_\theta(x^{(i)}) \right) + \alpha_0 || D f_\theta (x^{(i)})_0 ||^2_2 + \alpha_1 || D f_\theta (x^{(i)})_1 ||^2_2
\end{equation*}

where $\alpha_0$ and $\alpha_1$ are the regularization parameters for the different modes.

\begin{figure}[!htbp]
\centering
\includegraphics[width=1\linewidth]{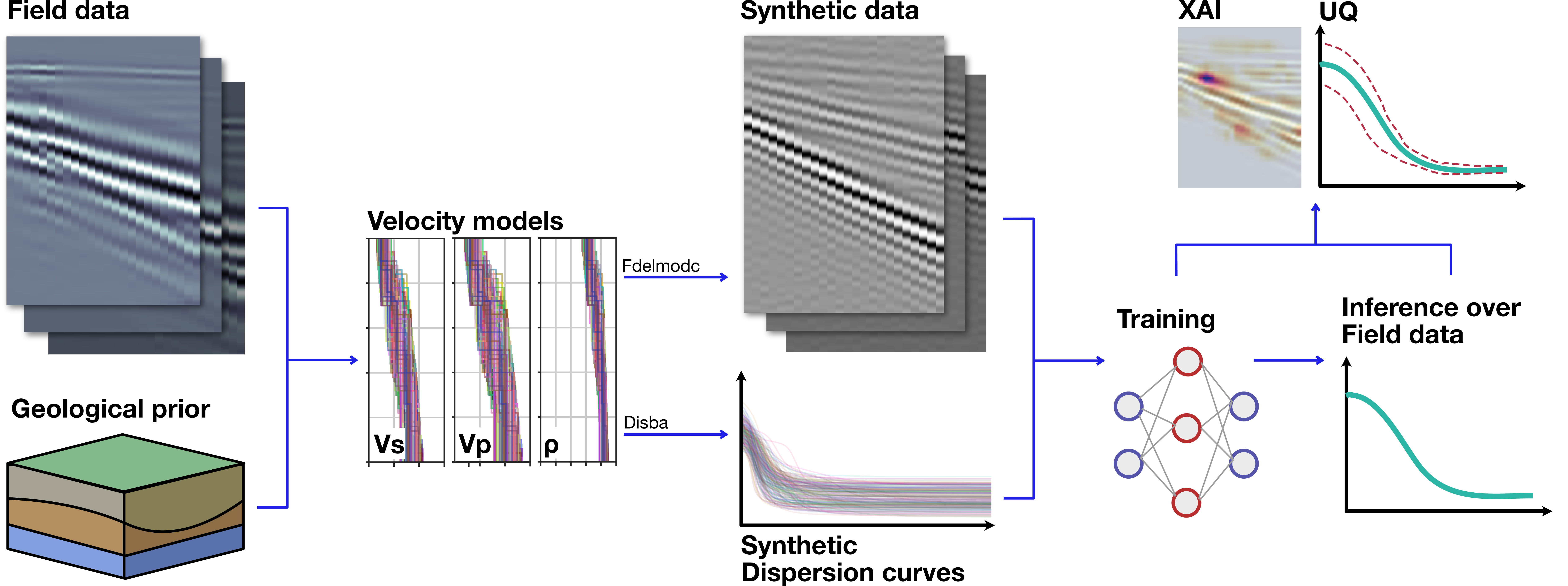}
\caption{A schematic representation of the proposed workflow.}
\label{fig1}
\end{figure}

\section{Numerical Results}

The proposed methodology is applied to a field dataset provided by Fugro under an academic license with the aim of automatically extracting dispersion curves after performing training purely on synthetic data. This dataset was acquired in 2018 for engineering purposes by firing 129 shots over a total length of 300 meters; each shot gather is composed of 24 traces with a receiver interval equal to one meter and an initial offset of 15 meters. The seismic response of the medium is recorded using a 0.5 ms sampling interval for a total recording time of one second. To begin with, a synthetic dataset is created that closely resembles the characteristics of the field data, following the steps described in the previous section. Special attention in the preparation of the training dataset is fundamental to ensure that the neural network can learn features from the synthetic data that are also representative for the field data at inference stage. The training dataset consists of 1000 velocity and density profiles, and their associated dispersion curves are shown in Figure 2. Finally, training is performed using a single NVIDIA GeForce GTX 1660 Ti GPU for a duration of around one hour.

\begin{figure}[!htbp]
\centering
\includegraphics[width=1\linewidth]{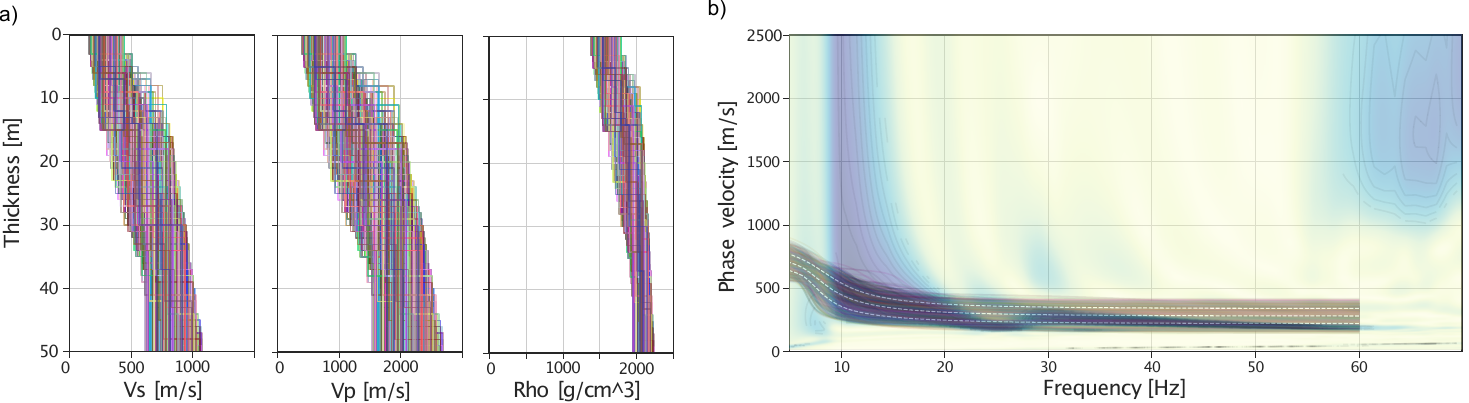}
\caption{a) S-wave, P-wave velocity, and density profiles for the 1000 randomly generated models, and b) associated dispersion curves displayed on top of the average dispersion spectra obtained from the field dataset. Note how the overall distribution of the synthetic dispersion curves nicely overlays the peaks associated with the fundamental and first modes of the average dispersion spectrum.}
\label{fig2}
\end{figure}

\subsection*{Inference}

Once trained, the neural network was applied on both synthetic and field data; for the synthetic case, a set of previously unseen shot gathers was produced to avoid bias in the prediction. While for the field dataset, a number of representative shot gathers were evaluated after applying the pre-processing steps detailed in the previous section. Figure 3 illustrates the obtained results. The synthetic data results exhibit a generally good fit between the predicted and true dispersion curves. Meanwhile, for the field dataset, a set of three gathers were selected based on their estimated visual similarity to the synthetic data. The shot gather in the first row was chosen for its high resemblance, and the inference for this input demonstrates a strong correlation between the dispersion spectra distribution and the expected behavior for the fundamental mode dispersion curve. The second row displays a gather that slightly out of the distribution; however, the network is still able to produce a satisfactory result for the fundamental mode. Finally, the shot gather in the last row is considered to be completely out of the distribution, as it exhibits wave dispersion phenomena that difficult to replicate in the synthetic dataset; nevertheless, the inference of our model displays a good resemblance with the maximum energy in the spectra while still maintaining an appropriate behavior in terms of smoothness and velocity range. It is important to note that the dispersion spectra shown in Figure 3 are used only for visualization purposes and to aid our assessment of the model performance, but they were not given to the model during the training or testing stages. 

\begin{figure}[!htbp]
\centering
\includegraphics[width=1\linewidth]{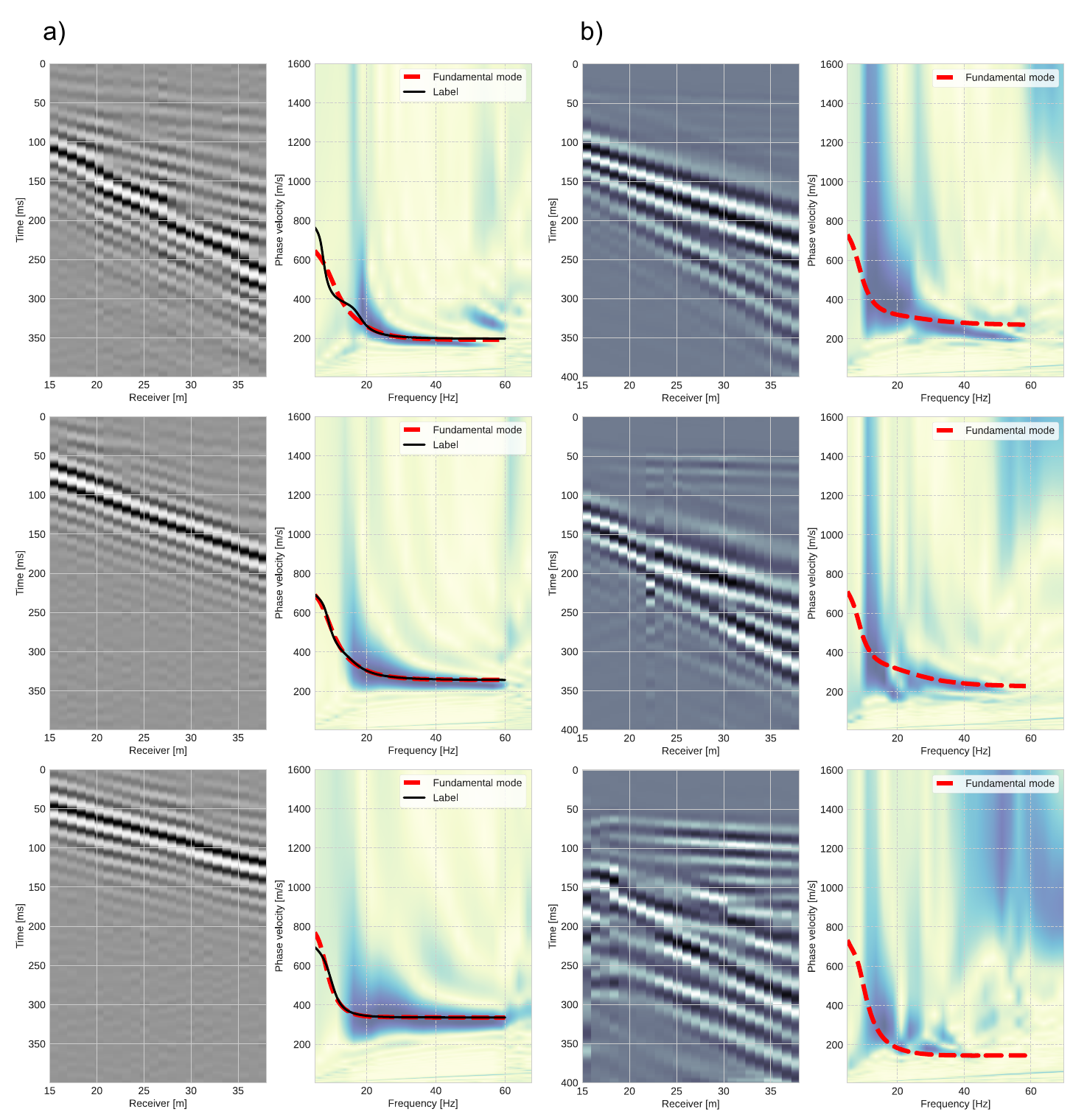}
\caption{Inference of fundamental mode performed over a set of a) synthetic shot gathers and b) field shot gathers. Dispersion spectra panels are only shown for visualization purposes.}
\label{fig3}
\end{figure}

\subsection*{Uncertainty Quantification}

In Machine Learning, Uncertainty Quantification (UQ) refers to the process of quantifying the uncertainty of a model's predictions. It is commonly used to identify the sources of error in a model's predictions, and to determine if the model has been successfully trained such that it can generalize well to unseen data (arising from the same distribution as the training samples). UQ can also be used to identify how sensitive a model’s prediction is to changes in the input data \cite{Abdar2021}. In this work, we have implemented two techniques in order to obtain insights into the quality of the predicted dispersion curves as a whole, as well as to produce frequency-dependent confidence levels. First, a Monte-Carlo dropout approach \cite{Gal2016} is introduced by randomly disabling neurons during the inference stage; by feeding the network multiple times with the same input, a set of outputs is generated from which we can compute sample statistics (e.g., mean, standard deviation). Figure 4 exhibits the obtained results for both synthetic and field gathers. We found the results of the Monte-Carlo approach to be overly smoothed and not particularly sensitive to the data used as input, therefore a different approach is also tested. For this, we employed the Distributional Parameter Estimation (DPE) methodology, in which instead of training the network to predict the fundamental mode dispersion curve (and possibly also the first mode dispersion curve), the network is trained to predict the distributional parameters of the output space by assuming that the output is normally distributed with parameters $\mu$ and $\sigma^2$.  Considering the case where we want to predict the fundamental mode only, this is implemented using a maximum likelihood approach, which entails minimizing the following loss function:

\begin{equation*}
\underset{\theta}{\mathrm{argmin}} \frac{1}{N_s} \sum^{N_s}_{i=1} \frac{|| y^{(i)} - \mu^{(i)} ||^2_2 }{2 \sigma^{2(i)}} + \frac{log \sigma^{(i)}}{2}
\end{equation*}

where $f_\theta\left(x^{(i)}\right)=[\mu^{\left(i\right)}, \sigma^{\left(i\right)}]$. Note that the regularization terms can also be included, playing the role in this case of prior on the mean parameter $\mu^{\left(i\right)}$.
 The estimated uncertainties for this second approach show a higher sensitivity to the input and an overall more accurate understanding of the estimated uncertainty in the predictions, as seen in the third column of Figure 4, for both synthetic and field inputs. We can see how the uncertainty is higher for the lower ranges of frequencies, while it decreases as the frequency increases. This is understandable since the variability of the dispersion curves is higher at lower frequencies, while at the same time, it flattens over a smaller range of possible values for the higher ranges of frequency. When compared to the synthetic gather, the field input appears to produce a slightly wider uncertainty over the same range of frequencies since the field gather contains seismic phenomena that could not be replicated in our synthetic dataset, and therefore represents a harder task for our model to predict.

\begin{figure}[!htbp]
\centering
\includegraphics[width=1\linewidth]{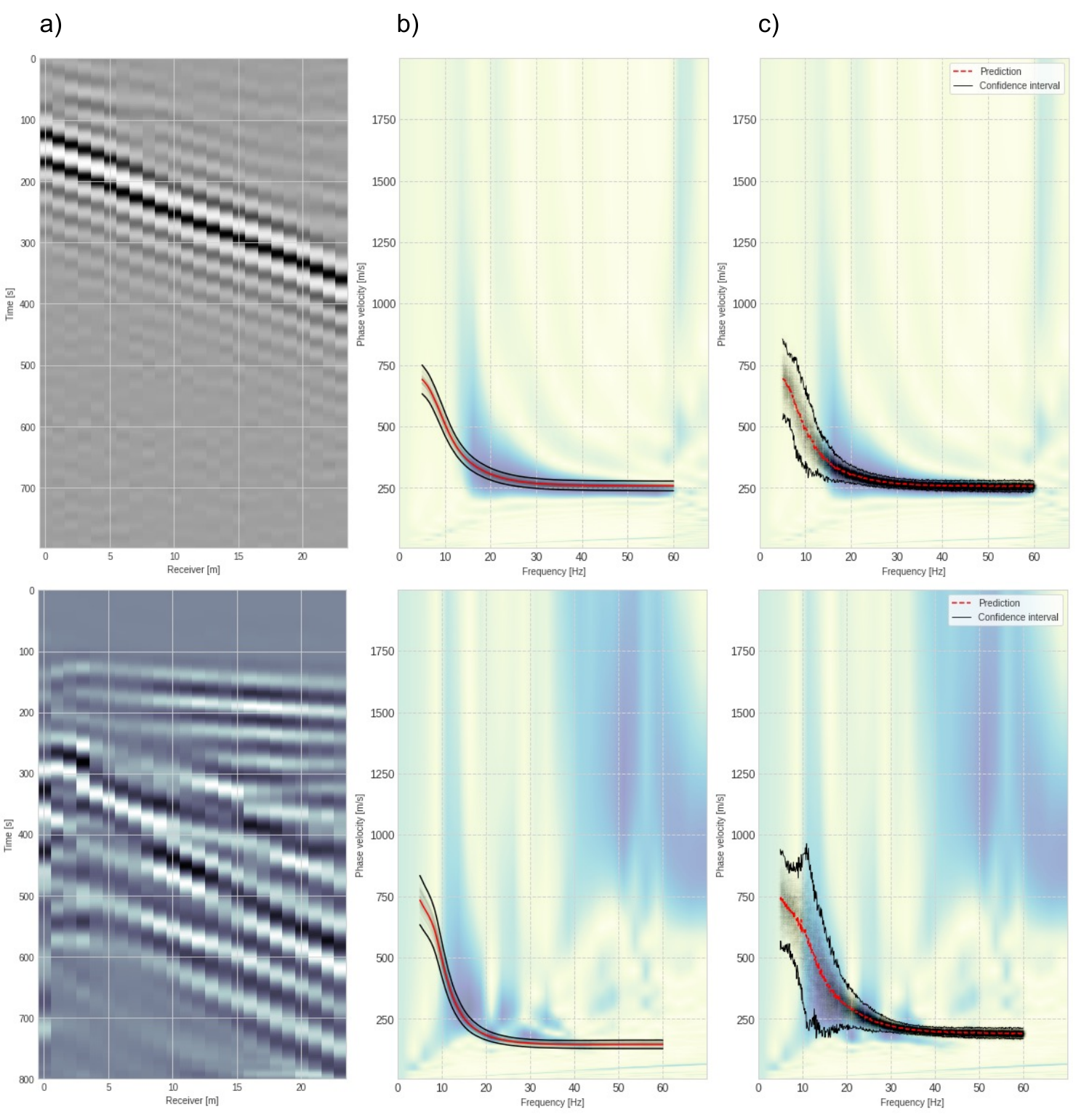}
\caption{a) Synthetic (first row) and field data (second row) shot gathers and their corresponding uncertainty quantification using b) Monte Carlo dropout approach and c) DPE approach.}
\label{fig4}
\end{figure}

\subsection*{CNN Visualization}

Explainable AI (referred to as XAI) is an emerging field in Machine Learning that aims at providing tools to interpret the decision making process of a neural network, in our case a CNN. Feature attribution is one family of such techniques, whereby the individual predictions of a network are explained by attributing each input feature according to how much it changes the prediction (positively, negatively, or with no influence). They can be classified into two groups: occlusion/perturbation-based, where the algorithm manipulates parts of the image to generate explanations in a model-agnostic way; or gradient-based, where the algorithm computes the gradient of the prediction with respect to the input features. The final output of both methods is a heatmap that quantifies each pixel's relevance to the prediction \cite{Christoph2022}. Occlusion \cite{Zeiler2013} is a pixel attribution method, representing a particular case of feature attribution methods specifically designed for images. This model interpretability technique has been explored to assess the focus areas in the input shot gathers of our trained network. The utilization of occlusion maps has demonstrated their efficacy and dependability in enhancing the comprehension of the model's behavior. As depicted in Figure 5, it is evident that the network predominantly concentrates on the dispersive constituents of the input shot gathers. Given that occlusion is calculated for each individual sample of the output, and considering the output of the neural network as a vector comprising 500 points, a total of 500 occlusion maps are generated, each corresponding to a distinct position along the frequency axis. A close examination of the associated figure reveals three separate frequency steps where the occlusion illustrates the network’s attention.

\begin{figure}[!htbp]
\centering
\includegraphics[width=1\linewidth]{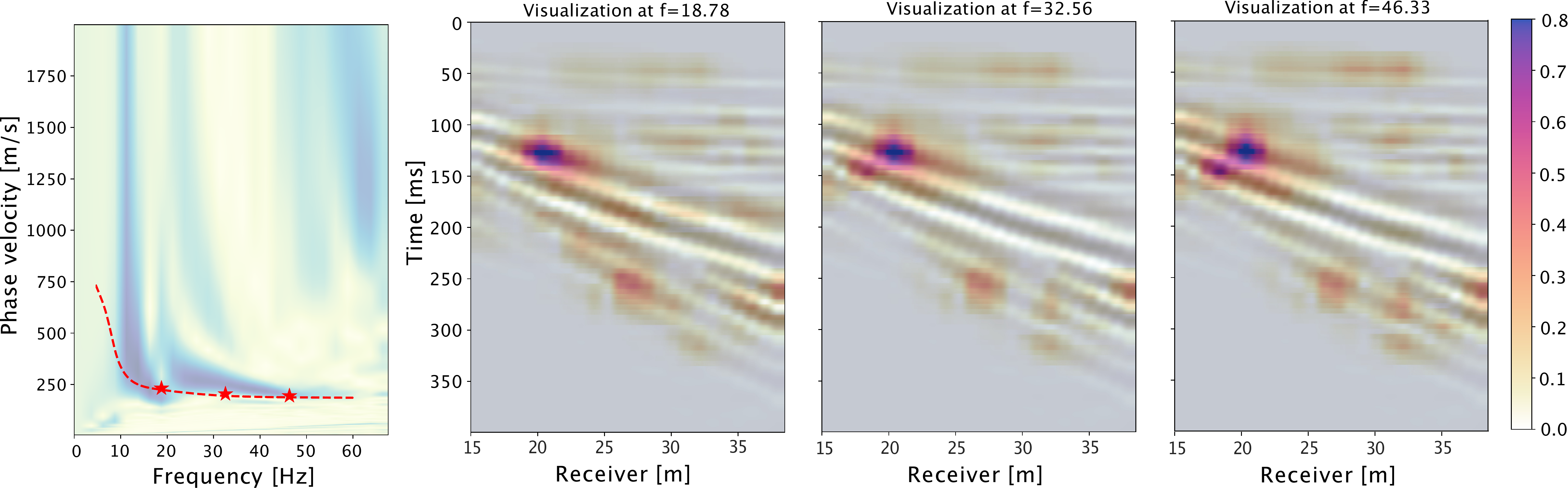}
\caption{Occlusion applied to a sample field shot gather. The first panel shows the model's output on top of the dispersion spectrum. Every other panel shows a heatmap of the importance of each pixel in the shot gather to the inference at the specified frequency (identified by the 3 stars in the first panel).}
\label{fig5}
\end{figure}

\section*{Inversion results}

After applying the proposed workflow to the entire field dataset, the retrieved phase velocity dispersion curves are inverted for a set of 1D isotropic layered velocity models using the Competitive Particle Swarm Optimization algorithm provided in the open-source library \textit{evodcinv} \cite{Luu2021}. As shown in Figure 6, the overall structure of the near subsurface is well recovered, and the individual inversion results are continuous amongst the horizontal definition of layers. For a selected profile, the inversion completed with a data misfit value of around 0.02. 

\begin{figure}[!htbp]
\centering
\includegraphics[width=1\linewidth]{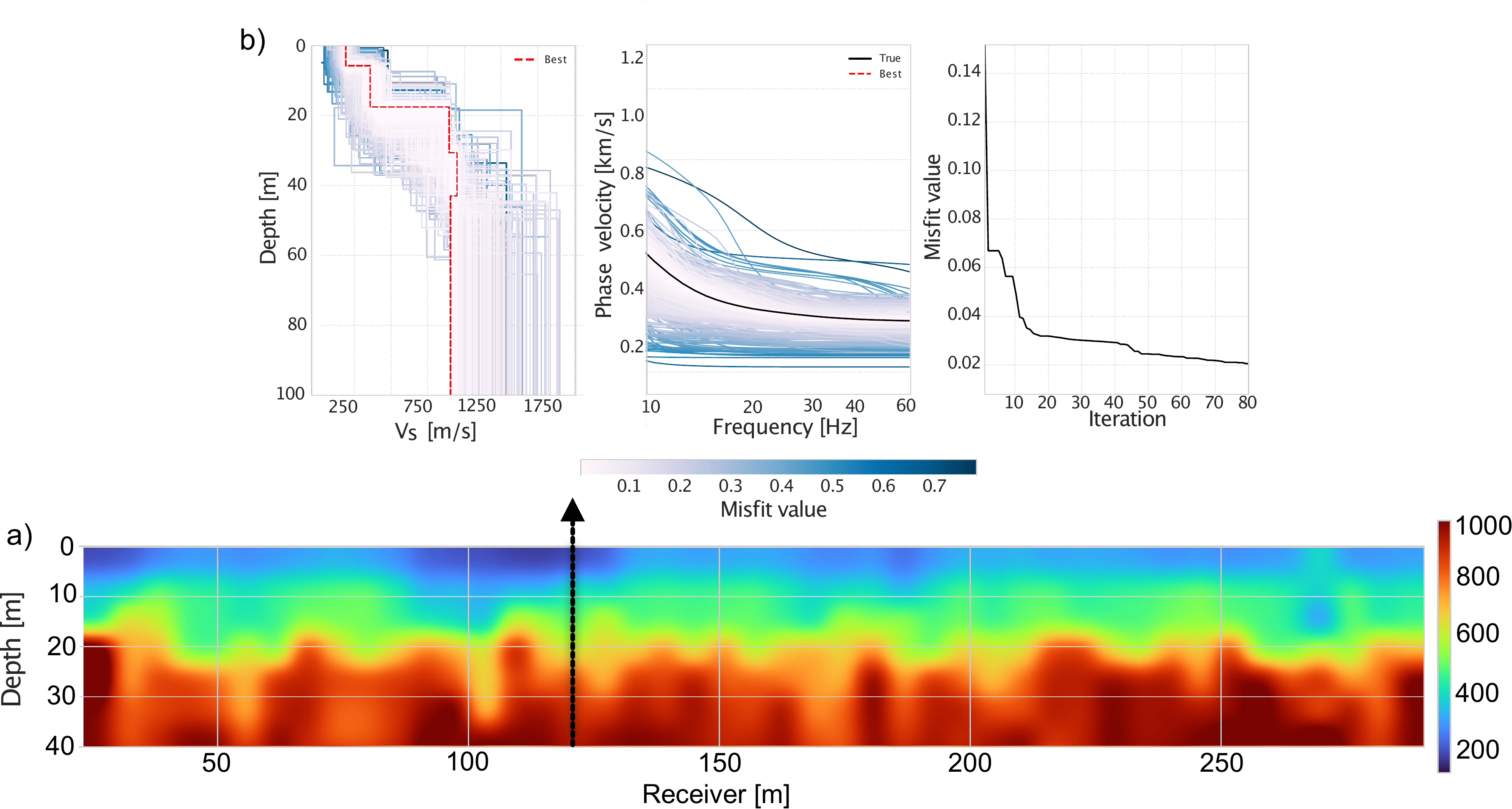}
\caption{a) 2D Inversion of the dispersion curves obtained by applying the trained model on all shot gathers of the field dataset. b) Inverted profiles, predicted data and data misfit as function of iterations for a sample dispersion curve at receiver 148.}
\label{fig6}
\end{figure}

\subsection*{Results of first mode implementation}

Finally, our neural network is re-trained to predict both the fundamental and first modes. Once trained, this model provides accurate predictions on the validation synthetic dataset, which are on par with the results shown for the fundamental mode only model. Predictions for the same shot gathers of Figure 3 are shown in Figure 7.  Generally, the neural network's predicted dispersion curves match those expected, as determined by both our labels and the dispersion spectra. As for the field data, we found that the predictions for the fundamental mode are not drastically different from the results of the previous iteration; similarly, the first mode prediction matches the dispersion spectra in most cases. We also observe that in the cases where the shot gathers are visually different from those used in the training dataset (e.g., shot gathers with poorly coupled receivers – see bottom row of Figures 3b and 7b), the introduction of the first mode appears to help the neural network learn and predict with more accuracy, given that the new task involves learning more features than before. These results are promising since, for a human interpreter, picking higher modes is already complex and hard to automatize with most of the previously described state-of-the-art techniques. 

\begin{figure}[!htbp]
\centering
\includegraphics[width=1\linewidth]{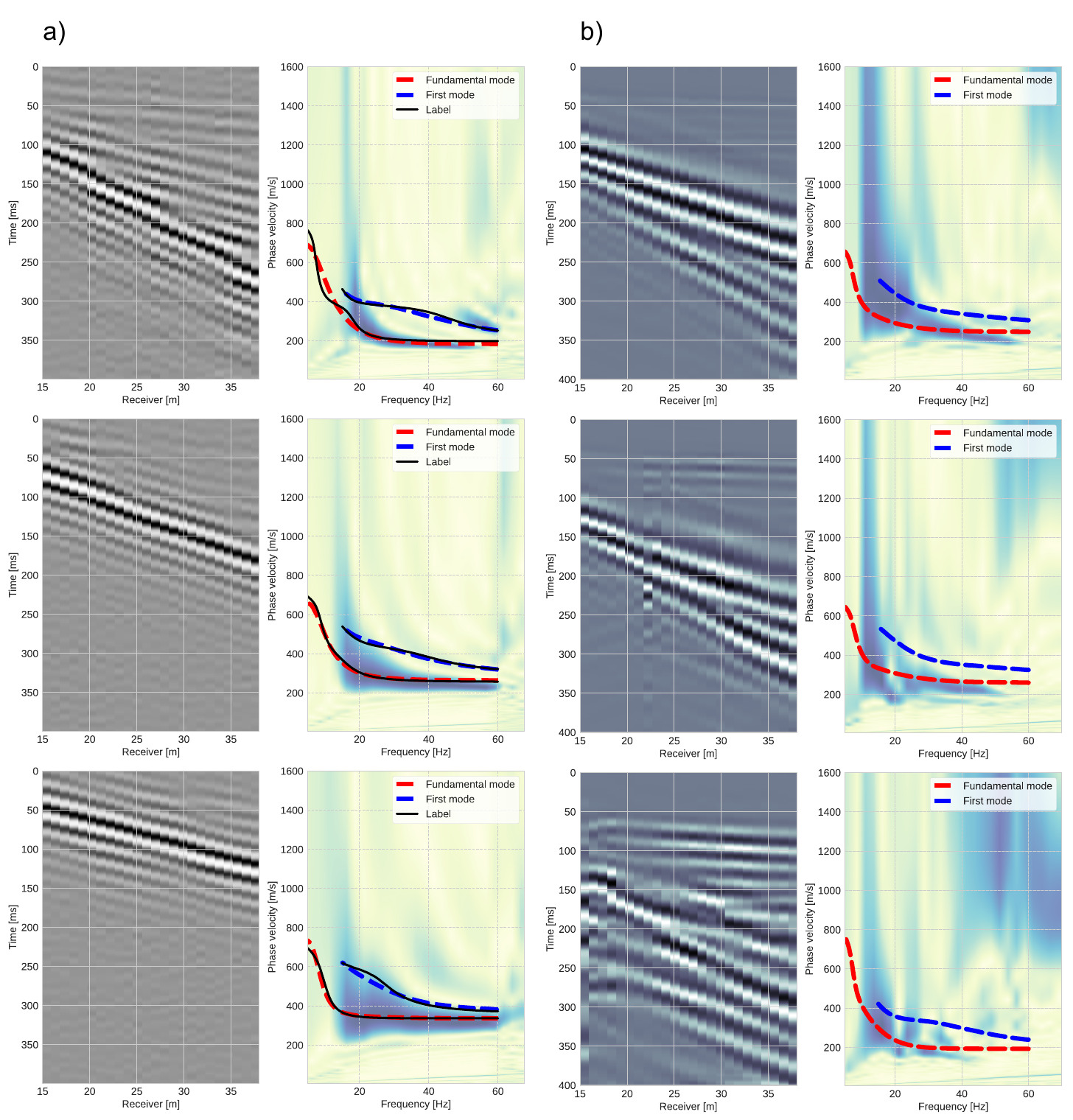}
\caption{Inference of fundamental and first modes performed over a set of a) synthetic shot gathers and b) field shot gathers. Dispersion spectra are only shown for visualization purposes.}
\label{fig7}
\end{figure}

Figure 8 displays the results of applying the inversion to the newly extracted dispersion curves. The inversion algorithm attempts to match both the fundamental and first modes, resulting in a more even and horizontally consistent model that better reflects the anticipated geology. Furthermore, the occurrence of artifacts, such as high-velocity zones arranged vertically along the profile, is reduced. We can see how the inclusion of the first mode enhances the quality of the inverted model, as expected, since it provides additional information to the inversion procedure that, in turn, constrains the results to be more on par with the anticipated distribution of velocities. 

\begin{figure}[!htbp]
\centering
\includegraphics[width=1\linewidth]{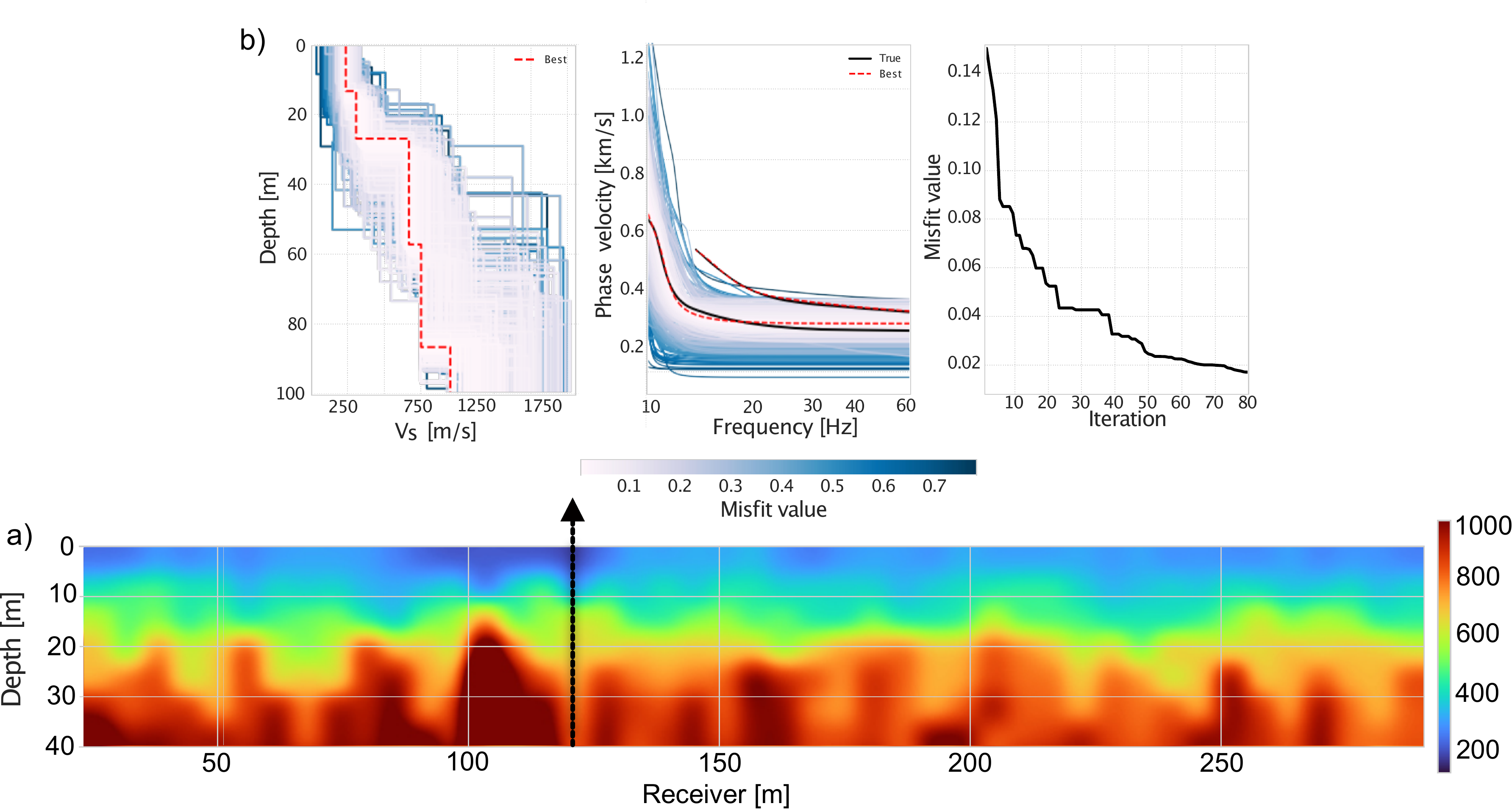}
\caption{a) 2D Inversion of the multi-mode dispersion curves obtained by applying the trained model on all shot gathers of the field dataset. b) inverted profile, estimated dispersion curve and data misfit compared with the number of iterations for a sample dispersion curve at receiver 148.}
\label{fig8}
\end{figure}

\section*{Conclusions}

In order to accelerate the process of extracting dispersion curves from seismic data to be used in the MASW workflow, this paper proposes a deep-learning-based methodology that directly maps seismic shot gathers to dispersion curves. Our methodology relies on the creation of realistic 1-D velocity and density profiles based on prior geological knowledge of the area of interest and the creation of synthetic seismic datasets using a full-wavefield modeling engine, e.g., elastic finite-difference. Experiments carried out on a synthetic dataset reveal that the proposed neural network has excellent performance, fast prediction time, and high accuracy. Several challenges arise during the application of the trained model to a field dataset: to improve the generalization capabilities of the model, various data pre-processing strategies are suggested, of which include: (1) the convolution of both the synthetic and field data with the wavelet extracted from the opposite dataset, and (2) the addition of colored noise and linear events to the synthetic dataset since these features are prominent in the field dataset. Finally, by inverting the extracted dispersion curves for an S-wave velocity model, we prove that the deep learning based dispersion curves can provide subsurface models consistent with previous knowledge of the area. We also show that the proposed approach can be easily extended to jointly estimate multiple modes: with the introduction of the first mode significantly enhancing the predicted dispersion curves for the field data, ultimately providing higher accuracy and a better-constrained inversion. We believe, our work lays the foundations for a fully automated MASW workflow. 

\section*{Acknowledgements}
The authors thank KAUST for supporting this research. Danilo Chamorro and Jiahua Zhao performed this work as part of the Visiting Student Research Program (VSRP). We also thank Fugro for releasing the field dataset and permission to show the associated results.  The accompanying code to reproduce the results in this paper can be found at \url{https://github.com/DIG-Kaust/ML_Dispersion.git}.

\bibliographystyle{unsrt}  
\bibliography{references}

\begin{thebibliography}{10}

\bibitem{Moro2015}
Giancarlo~Dal Moro.
\newblock Surface wave analysis for near surface applications.
\newblock {\em Surface Wave Analysis for Near Surface Applications}, pages
  87--102, 2015.

\bibitem{Park2007}
Choon~B. Park, Richard~D. Miller, Jianghai Xia, and Julian Ivanov.
\newblock Multichannel analysis of surface waves (masw) - active and passive
  methods.
\newblock {\em Leading Edge (Tulsa, OK)}, 26:60--64, 4 2007.

\bibitem{Penumadu2005}
Dayakar Penumadu and Choon~B. Park.
\newblock Multichannel analysis of surface wave (masw) method for geotechnical
  site characterization.
\newblock {\em Leading Edge}, pages 1--10, 10 2005.

\bibitem{Socco2004}
L.V. Socco and C.~Strobbia.
\newblock Surface‐wave method for near‐surface characterization: a
  tutorial.
\newblock {\em Near Surface Geophysics}, 2:165--185, 8 2004.

\bibitem{Baglari2018}
Dipjyoti Baglari, Arindam Dey, and Jumrik Taipodia.
\newblock A state-of-the-art review of passive masw survey for subsurface
  profiling.
\newblock {\em Innovative Infrastructure Solutions}, 3, 12 2018.

\bibitem{Zheng2014}
D.~Zheng and X.~G. Miao.
\newblock Multimodal rayleigh wave dispersion curve picking and inversion to
  build near surface shear wave velocity models.
\newblock {\em 76th EAGE Conference and Exhibition 2014, Workshops}, pages
  175--177, 6 2014.

\bibitem{Taipodia2020}
J.~Taipodia, A.~Dey, S.~Gaj, and D.~Baglari.
\newblock Quantification of the resolution of dispersion image in active masw
  survey and automated extraction of dispersion curve.
\newblock {\em Computers \& Geosciences}, 135:104360, 2 2020.

\bibitem{Colombo2019}
Taqi~Yousuf Alyousuf, Daniele Colombo, and Saudi Aramco.
\newblock Advances in surface-wave analysis using single sensor seismic data
  and deep neural network algorithm for near surface characterization.
\newblock {\em Leading Edge}, 2019.

\bibitem{Rovetta2020}
D.~Rovetta, A.~Kontakis, and D.~Colombo.
\newblock Fully automatic picking of surface wave dispersion curves through
  density-based spatial clustering.
\newblock {\em Leading Edge}, 2020:1--5, 12 2020.

\bibitem{Kaul2021}
A.~Kaul, P.J. Bilsby, A.~Misbah, and A.~Abubakar.
\newblock Machine-learning-driven dispersion curve picking for surface-wave
  analysis, modelling, and inversion.
\newblock {\em 82nd EAGE Annual Conference \& Exhibition}, 2021:1--5, 10 2021.

\bibitem{Pan2019}
Yudi Pan, Lingli Gao, and Thomas Bohlen.
\newblock High-resolution characterization of near-surface structures by
  surface-wave inversions: From dispersion curve to full waveform.
\newblock {\em Surveys in Geophysics}, 40:167--195, 3 2019.

\bibitem{Boaga2013}
Jacopo Boaga, Giorgio Cassiani, Claudio~L. Strobbia, and Giulio Vignoli.
\newblock Mode misidentification in rayleigh waves: Ellipticity as a cause and
  a cure.
\newblock {\em Geophysics}, 78, 6 2013.

\bibitem{Zhang2003}
Shuang~X. Zhang and Lung~S. Chan.
\newblock Possible effects of misidentified mode number on rayleigh wave
  inversion.
\newblock {\em Journal of Applied Geophysics}, 53:17--29, 2003.

\bibitem{Virieux2012}
J.~Virieux.
\newblock P-sv wave propagation in heterogeneous media: Velocity‐stress
  finite‐difference method.
\newblock {\em https://doi.org/10.1190/1.1442147}, 51:889--901, 2 2012.

\bibitem{Dunkin1965}
John~W. Dunkin.
\newblock Computation of modal solutions in layered, elastic media at high
  frequencies.
\newblock {\em Bulletin of the Seismological Society of America}, 55:335--358,
  4 1965.

\bibitem{Thorbecke2011}
Jan~W. Thorbecke and Deyan Draganov.
\newblock Finite-difference modeling experiments for seismic interferometry.
\newblock {\em Geophysics}, 76, 12 2011.

\bibitem{Luu2021}
Keurfon Luu.
\newblock evodcinv: Inversion of dispersion curves using evolutionary
  algorithms.
\newblock {\em None}, 2021.

\bibitem{He2015}
Kaiming He, Xiangyu Zhang, Shaoqing Ren, and Jian Sun.
\newblock Deep residual learning for image recognition.
\newblock {\em arXiv}, 12 2015.

\bibitem{Abdar2021}
Moloud Abdar, Farhad Pourpanah, Sadiq Hussain, Dana Rezazadegan, Li~Liu,
  Mohammad Ghavamzadeh, Paul Fieguth, Xiaochun Cao, Abbas Khosravi, U.~Rajendra
  Acharya, Vladimir Makarenkov, and Saeid Nahavandi.
\newblock A review of uncertainty quantification in deep learning: Techniques,
  applications and challenges.
\newblock {\em Information Fusion}, 76:243--297, 12 2021.

\bibitem{Gal2016}
Yarin Gal and Zoubin Ghahramani.
\newblock Dropout as a bayesian approximation: Representing model uncertainty
  in deep learning zoubin ghahramani.
\newblock {\em arXiv}, 2016.

\bibitem{Christoph2022}
Molnar Christoph.
\newblock {\em Interpretable Machine Learning: A Guide for Making Black Box
  Models Explainable}.
\newblock None, 2022.

\bibitem{Zeiler2013}
Matthew~D. Zeiler and Rob Fergus.
\newblock Visualizing and understanding convolutional networks.
\newblock {\em Lecture Notes in Computer Science (including subseries Lecture
  Notes in Artificial Intelligence and Lecture Notes in Bioinformatics)}, 8689
  LNCS:818--833, 11 2013.

\end{thebibliography}

\end{document}